\documentclass[lettersize,journal]{IEEEtran}

\usepackage{amsmath,amsfonts}
\usepackage{algorithmic}
\usepackage{algorithm}
\usepackage{array}
\usepackage{textcomp}
\usepackage{stfloats}
\usepackage{url}
\usepackage{verbatim}
\usepackage{graphicx}
\usepackage{cite}
\usepackage{setspace}
\usepackage{enumitem}
\usepackage{stfloats}
\usepackage{bbm}
\usepackage{amsthm,amssymb}
\usepackage{balance}
\usepackage{subfigure}
\usepackage{subcaption}
\usepackage{color}
\usepackage{booktabs} 
\usepackage{threeparttable}
\usepackage{xcolor}
\usepackage{tabularx,colortbl}
\usepackage{multirow}
\usepackage{makecell}
\usepackage{ifthen}
\usepackage{ulem}
\usepackage{caption}
\usepackage[caption=false,font=normalsize,labelfont=sf,textfont=sf]{subfig}

\newcommand\bflag[1]{\textcolor{black}{#1}}

\newif\ifshowdeleted
\showdeletedtrue

\newcommand{\vecx}{\mathbf{x}}

\usepackage{mathrsfs}

\begin{document}

\title{Rapid LoRA Aggregation for Wireless Channel Adaptation in Open-Set Radio Frequency Fingerprinting}

\author{Mingxi Zhang, Renjie Xie,~\IEEEmembership{Member,~IEEE}, Jincheng Wang, Guyue Li,~\IEEEmembership{Member,~IEEE}, Wei Xu,~\IEEEmembership{Fellow,~IEEE}
        %
\thanks{M. Zhang, R. Xie, and J. Wang are with the School of Internet of Things, Nanjing University of Posts and Telecommunications, Nanjing 210003, China~(e-mails: B23080111@njupt.edu.cn, renjie\_xie@njupt.edu.cn, B24080122@njupt.edu.cn)}%
\thanks{Guyue Li is with the School of Cyber Science and Engineering, Southeast University, Nanjing 210096, China~(e-mail: guyuelee@seu.edu.cn).}
\thanks{W. Xu is with the National Mobile Communications Research Lab, Southeast University, Nanjing 210096, China (e-mail: wxu@seu.edu.cn).}}

\maketitle

\begin{abstract}
Radio frequency fingerprints~(RFFs) enable secure wireless authentication but struggle in open-set scenarios with unknown devices and varying channels. Existing methods face challenges in generalization and incur high computational costs. We propose a lightweight, self-adaptive RFF extraction framework using Low-Rank Adaptation~(LoRA). By pretraining LoRA modules per environment, our method enables fast adaptation to unseen channel conditions without full retraining. During inference, a weighted combination of LoRAs dynamically enhances feature extraction. Experimental results demonstrate a 15\% reduction in equal error rate (EER) compared to non-fine-tuned baselines and an 83\% decrease in training time relative to full fine-tuning, using the same training dataset. This approach provides a scalable and efficient solution for open-set RFF authentication in dynamic wireless vehicular networks. 
\end{abstract}


\section{Introduction}
\IEEEPARstart{R}{adio} frequency fingerprints~(RFFs) are unique physical-layer characteristics inherent to specific wireless devices. These fingerprints emanate naturally during the manufacturing process and are challenging to alter or manipulate~\cite{wang2016wireless}. Leveraging RFFs for authentication offers significant benefits over traditional higher-layer methods, including reduced power consumption and minimized latency\cite{hou2014physical}. With the rapid development of Internet of Things~(IoT) applications, the rising number of malicious attacks threatens information security and system stability. In this context, ensuring wireless network security has become increasingly critical and employing RFF for authentication could be a promising solution~\cite{xie2024radio, jagannath2022comprehensive}.

As RFFs are embedded within the transmitted signals which propagate through a wireless channel before reaching the receivers~\cite{danev2012physical}
, utilizing RFF for authentication requires extracting distinctive RFF features from the received signals. Recently, convolutional neural networks (CNNs)~\cite{zhang2024DFLNet, shen2024federated,del2024fingerprint,iyiparlakouglu2024optimizing} have shown great promise for RFF extraction. 
These approaches leverage deep neural networks to capture nonlinear characteristics from preprocessed received signals and use these distinguishing features as the basis for device classification.


From another perspective, the aforementioned methods regarded RFF authentication as a closed-set classification problem since all test devices were included in the training dataset. However, we could only collect a small amount of relevant data in real-world scenarios due to practical time and labor costs. Consequently, these methods failed to identify unknown devices accurately. Moreover, signal receivers are likely to encounter numerous attacks from different unknown devices~\cite{xie2024radio}. Therefore, the actual situation requires a reliable RFF identifier capable of distinguishing both known and unknown devices. This constitutes the well-known open-set physical-layer authentication problem~\cite{xie2021generalizable, hanna2021open, zhang2024real}.
To address the open-set problem, extracted RFF features need to be highly separable. However, preprocessing procedures in previous works~\cite{merchant2018deep,peng2023supervised} may discard important device-related identity information, thereby reducing feature separability. Based on this circumstance, \cite{xie2021generalizable} proposed a model-and-data driven preprocessing module and adapted maximum likelihood estimation~(MLE) to the metric learning framework to ensure high separability of the extracted RFF features. However, the reliance on MLE would easily leads to overfit to the training data and loses the ability to generalize to different propagation environments.

To enhance the generalizability of the RFF features and reduce the impact of varying propagation environments, prior studies often adopted data augmentation (DA) techniques, such as those introduced in \cite{yu2019robust} and \cite{shen2022towards}. However, these approaches may either discard critical information or preserve irrelevant features, as the extracted representations are designed to be robust under artificially generated channel conditions that may not adequately reflect real-world scenarios. To address this limitation, \cite{xie2023disentangled} proposed a self-supervised disentangled representation (DR) framework. By leveraging intrinsic variations in propagation environments present within the training data, this method enables more realistic data augmentation and improves robustness in unseen environments. Nevertheless, it entails high computational costs during training and depends on the assumption that shuffling operations can generate sufficient synthetic samples to approximate the distribution of unknown signals. This assumption necessitates a diverse training dataset. When such diversity is insufficient, 
performance on open-set channel conditions may fall short of expectations. 

To address this challenge, it is crucial to enhance the adaptability of the RFF extractor to varying propagation environments, such as high-mobility conditions encountered in vehicular systems. In this work, we introduce a computationally efficient and cost-effective adaptation framework based on Low-Rank Adaptation (LoRA)~\cite{hu2022lora}, specifically tailored for the RFF extractor to facilitate rapid generalization to unseen devices across diverse wireless channel conditions. Compared with non-fine-tuning approaches such as DR-RFF, the proposed method demonstrates significantly enhanced adaptability in complex environments. Moreover, unlike conventional fine-tuning strategies, including full fine-tuning\cite{shen2024federated, huan2025LLRF} and standard LoRA, it achieves superior performance while substantially lowering computational costs.

\section{System Overview and Problem Statement}
This section presents the open-set RFF authentication scenario and discusses the associated performance degradation due to varying channel conditions.

\subsection{Open-set RFF Authentication}
Consider a system comprising multiple transmitting terminals and a single receiver. Let the preamble signal of length $M$ be represented by $\mathbf{s} \in \mathbb{C}^{M}$. The transmitted signal is first distorted by the unique hardware impairments of the transmitter, then further modified during propagation through the wireless channel, and ultimately received as $\mathbf{x} \in \mathbb{C}^{M}$. This end-to-end signal transmission process can be described by the model:
\begin{equation}
\mathbf{x} = g(f(\mathbf{s})),
\end{equation}
where $f: \mathbb{C}^{M} \to \mathbb{C}^{M}$ accounts for the transmitter-specific hardware impairments, and $g: \mathbb{C}^{M} \to \mathbb{C}^{M}$ models the effects of the propagation channel.

A typical RFF authentication system employs an RFF extractor $ F: \mathbb{C}^{M} \to \mathbb{R}^{d} $ to extract unique and inherent hardware-specific features from the received signal $ \mathbf{x} $. This process is mathematically represented as:
\begin{equation}
    \mathbf{z} = F(\mathbf{x}),
\end{equation}
where $F$ is typically implemented as a deep neural network trained on a dataset $\mathcal{D} = \{(\mathbf{x}_i, \mathbf{y}_i)_G\}_{i=1}^{N}$ collected from $J$ known transmitters $\{\text{Tx}_1, \text{Tx}_2, \ldots, \text{Tx}_J\}$. The label $\mathbf{y}_i$ is a one-hot vector with $1$ at position $y$ and zeros elsewhere, where $y \in \mathbb{N}$ indicates the index of the transmitting device $\text{Tx}_y$. The notation $G = \{g_1, g_2, \ldots, g_L\}$ denotes the set of $L$ propagation environments associated with the signal samples.

Typically, open-set authentication based on metric learning employs cosine distance, defined as $D_\text{cos}(\mathbf{a}, \mathbf{b}) = 1 - \cos(\mathbf{a}, \mathbf{b})$, to assess whether two extracted RFFs,$\mathbf{z}_i$ and $\mathbf{z}_j$, belong to the same device~\cite{xie2021generalizable}. Specifically,
\begin{equation}
    \left\{
    \begin{array}{ll}
        D_\text{cos}(\mathbf{z}_i; \mathbf{z}_j) \leq T & \Rightarrow \quad \text{Same device}, \\
        D_\text{cos}(\mathbf{z}_i; \mathbf{z}_j) > T & \Rightarrow \quad \text{Different devices},
    \end{array}
    \right.
    \label{eq:measure}
\end{equation}
where $ T $ is a predefined threshold.

To achieve satisfactory identification performance, the extracted RFF $\mathbf{z}$ must be highly discriminative and predominantly reflect the hardware-specific imperfections captured by $f(\cdot)$. Given an auxiliary linear classifier $\mathbf{W} = \{\mathbf{w}_j\}_{j=1}^J$ and a scaling factor $\delta > 0$, the metric learning framework can be formulated as a maximum likelihood estimation (MLE) problem~\cite{xie2021generalizable}, expressed as
\begin{equation}
\begin{aligned}
\min_{F} \mathcal{L}(F, \mathcal{D}) = -\mathbb{E}_{(\mathbf{x}, y) \in \mathcal{D}}\left[\ln p(y\vert \mathbf{x})\right],
\label{eq:MLE}
\end{aligned}
\end{equation}
where 
\begin{equation}
\begin{aligned}
p(y\vert \mathbf{x}) = \frac{e^{\delta \cdot \cos(\mathbf{w}_y, \mathbf{z})}}{\sum_{j=1}^J e^{\delta \cdot \cos(\mathbf{w}_j, \mathbf{z})}}.
\end{aligned}
\end{equation}
Under this formulation, open-set RFF extraction can be effectively modeled as a conventional softmax-based metric learning task.

\subsection{Performance Degradation Caused by Channel Shifts}
Due to environmental limitations and practical constraints related to time and labor costs, the training set $\mathcal{D}$ typically captures only a limited range of channel conditions. In real-world deployment scenarios, wireless channels are highly dynamic and complex, often diverging significantly from those represented in the training data. This mismatch can result in overfitting to specific channel characteristics observed during training, thereby undermining the model's ability to generalize.

To address performance degradation arising from varying channel conditions, \cite{xie2023disentangled} proposes an adversarial regularization approach based on Disentangled Representation Learning (DRL). However, when the operational environments $G^\prime$ differ substantially from the training environment $G$, particularly in cases where $G \cap G^\prime = \emptyset$, the performance of RFF-based authentication systems may suffer significant deterioration. Under such extreme channel shifts, it becomes essential to enable rapid adaptation of the RFF extractor to new environments using minimal supervisory signals.

\section{Rapid LoRA Aggregation for Channel-Robust RFF Adaptation}
In this section, we first formally define the adaptation problem, then provide an overview of the fundamental principles of parameter-efficient fine-tuning (PEFT) ~\cite{hu2022lora}, and subsequently present a comprehensive description of the proposed RLA method.
\subsection{The Formulation of Channel Adaptation}
To ensure that the adaptive method can quickly adjust to unseen channel environments, we first establish a formal problem definition. In practical deployment scenarios, such as for new devices or previously unobserved channels, only a small dataset $\mathcal{D}^\prime = \{(\mathbf{x}_i^\prime, \mathbf{y}_i^\prime)_{G^\prime}\}_{i=1}^{N^\prime}$ is typically available, where $N' \ll N$. Given a pre-trained RFF feature extractor $F$, the goal is to learn an adapted model $F^\prime = F + \Delta F$ by solving the optimization problem:
\begin{equation}
\min_{\Delta F} \mathcal{L}(F + \Delta F, \mathcal{D}^\prime), 
\label{eq:delta}
\end{equation}
where $\mathcal{L}(\cdot)$ denotes the MLE loss defined in (\ref{eq:MLE}), and $\Delta F$ represents a parameter-efficient adapted module. The primary challenge involves achieving rapid and stable convergence despite limited training data and stringent computational constraints, both of which are essential for real-time authentication systems.

\begin{figure*}
    \centering
    \includegraphics[width=1\linewidth]{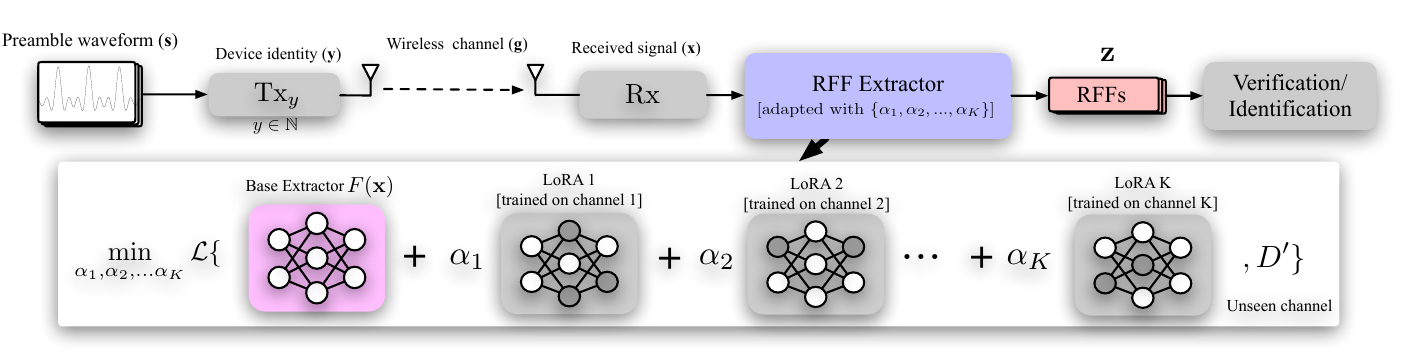}
    \caption{Rapid LoRA Aggregation for Wireless Channel Adaptation in Open-Set RFF Authentication.}
    \label{fig:overview}
\end{figure*}

\subsection{Parameter-efficient fine-tuning}
Traditional adaptation strategies often rely on full Fine-Tuning~(FT), which updates all parameters of the pre-trained model $F$ to minimize the loss on target data $\mathcal{D}^\prime$. While effective in principle, FT suffers from significant drawbacks in practical edge deployments. First, it requires storing and updating a large number of parameters, which leads to high memory consumption and computational cost. Second, FT is highly prone to overfitting when only limited labeled data $\mathcal{D}_{G^\prime}$ are available, especially under domain shifts such as those induced by varying wireless channel conditions. Finally, the full-parameter update strategy impedes model reuse and complicates version control in real-world systems where models must be frequently reloaded and adapted across diverse environments.

To overcome these limitations, recent advances have introduced PEFT methods that only updates a small set of parameters while keeping the pre-trained weights frozen. Among these, Low-Rank Adaptation~(LoRA) has emerged as a particularly compelling approach. The core idea of LoRA~\cite{hu2022lora} is to approximate the parameter update $\Delta F$ as a low-rank matrix decomposition, for all weight matrix:
\begin{equation}
\Delta \mathbf{W} = \mathbf{A}\mathbf{B}, \quad \text{where} \quad \mathbf{A} \in \mathbb{R}^{d_1 \times r}, \mathbf{B} \in \mathbb{R}^{r \times d_2},
\end{equation}
with $r \ll \min(d_1, d_2)$. Here, $\mathbf{W} \in \mathbb{R}^{d_1\times d_2}$ denotes a weight matrix in the original RFF feature extractor $F$. Instead of directly learning $\Delta \mathbf{W}$, LoRA learns two trainable matrices $\mathbf{A}$ and $\mathbf{B}$. The adapted layer output becomes:
\begin{equation}
\mathbf{h}' = (\mathbf{W} + \Delta\mathbf{W})\mathbf{x} = \mathbf{W}\mathbf{x} + \mathbf{A}(\mathbf{B}\mathbf{x}),
\end{equation}
where $\mathbf{x}$ is the input feature vector. Apparently, LoRA greatly reduced the number of trainable parameters compared with those in FT, thus noticeably reduces the computational overhead and memory usage. However, this modification also reduces the parameter space of the model and therefore weakens the learning ability of the model. Although we can address the overfitting problem in FT to some extent by correctly setting the value of $r$, the model may take more time to adapt and converge. These issues motivate the development of a more structured and agile adaptation framework which we present next.

\subsection{Rapid LoRA Aggregation for Channel Adaptation}
To achieve robust and efficient adaptation under channel variation, we propose Rapid LoRA Aggregation (RLA), a novel framework that leverages structured combinations of pre-trained low-rank modules to enable fast model specialization with minimal retraining.

As shown in Fig.\ref{fig:overview}, Formally, suppose we have a set of $K$ pre-trained LoRA modules $\{\Delta F_k\}_{k=1}^K$, each learned from data collected under a distinct channel condition $G_k$. These modules represent low-rank adjustments to the shared base model $F$, capturing environment-specific deviations in wireless propagation characteristics. Instead of fine-tuning a new LoRA from scratch for an unseen environment $G'$, RLA constructs the adaptation module $\Delta F'$ as a weighted aggregation of existing modules:
\begin{equation}
\Delta F' = \sum_{k=1}^{K} \alpha_k \cdot \Delta F_k,
\label{eq:loRA_agg}
\end{equation}
where the mixing coefficients $\alpha_k \in \mathbb{R}$ are the only trainable parameters.

This aggregation strategy offers three key benefits. First, it supports zero-shot or few-shot generalization by allowing the model to interpolate among known channel conditions based on similarity, even when no labeled data is available for $G'$. Second, it enables extremely fast adaptation: once the weights $\{\alpha_k\}$ are determined, applying the composite update requires only a lightweight forward pass without backpropagation.  Third, it significantly enhances memory and parameter efficiency by storing merely $K+1$ compact LoRA modules, i.e., the $K$ source adapters combined with a minimal fusion network, thereby substantially reducing both storage overhead and deployment costs.

Accordingly, the optimization problem in (\ref{eq:delta}) can be reformulated as
\begin{equation}
\min_{\alpha_1, \alpha_2,\dots,\alpha_K} \mathcal{L}(F + \Delta F', \mathcal{D}^\prime),
\label{eq:RLA}
\end{equation}
where the objective is to learn optimal aggregation weights using limited data from the target environment. 

To solve (\ref{eq:RLA}) efficiently without gradient computation, we employ the Covariance Matrix Adaptation Evolution Strategy (CMA-ES)~\cite{hansen2016cma}, a derivative-free optimization algorithm particularly effective for low-dimensional, non-convex problems. The method maintains a multivariate normal distribution over the coefficient vector $\boldsymbol{\alpha} = (\alpha_1, \dots, \alpha_K)$, parameterized by a mean vector $\boldsymbol{\mu}$ and covariance matrix $\mathbf{C}$. At each iteration, a population of candidate solutions $\{\boldsymbol{\alpha}^{(i)}\}_{i=1}^N$ is sampled from this distribution, balancing exploitation of high-performing regions with exploration through normally distributed perturbations.

\bflag{It is worth noting that the recently popular Mixture-of-Experts (MoE) \cite{shazeer2017moe} architecture also employs a multi-module structure to enhance model performance, but the method proposed in this paper differs fundamentally from MoE. First, MoE typically trains multiple modules as a whole and then activates them sparsely to obtain corresponding experts, whereas the proposed RLA combines individually trained LoRA modules to form a dense model for inference. Second, MoE often requires balancing module activation and relies on classical gradient descent for training, while RLA employs CMA-ES to optimize only a small set of coefficients, achieving faster fitting speeds. This makes it particularly suitable for resource-constrained IoT and vehicular applications with limited computational resources that demand high efficiency.}

\section{Experiment Evaluation}

In this section, we assess the effectiveness of the proposed RLA-based RFF adaptive method using real-world collected data. We evaluate and compare the performance and efficiency of RLA-RFF against established channel adaptation approaches, including full Fine-Tuning~(FT) in \cite{huan2025LLRF}, self-supervised pre-training~(SSL) followed by FT in \cite{shen2024federated}, and original LoRA~\cite{hu2022lora}. 

\begin{table}[!t]  
\caption{Dataset for Experiments}  
\centering  
\resizebox{\linewidth}{!}{	
\begin{threeparttable}[b]
\begin{tabular}{c|c|c}  
\toprule
Datasets & Devices IDs & Relationship between Txs and the Rx \\   
\hline
Training set & \multirow{2}*{1-45} & \multirow{2}*{separated by 0.3 to 1 meter} \\
\cline{1-1}
Validation set &  & \multirow{2}*{(LoS channel)} \\
\cline{1-2}
T1 & 46-54 &  \\  
\hline
T2 & 1-45 & separated by 0.3 to 1 meter \\  
\cline{1-2}
T3 & 46-54 &  (LoS, with device aging) \\ 
\hline
U1 & \multirow{3}*{55-59} & separated by one rooms~(NLoS)  \\  
\cline{1-1}
\cline{3-3}
U2 &  & separated by two rooms~(NLoS)  \\ 
\cline{1-1}
\cline{3-3}
U3 &  & separated by 40 meter~(Long LoS)  \\ 
\bottomrule
\end{tabular}  
\end{threeparttable}
}
\label{tb:dataset}
\end{table}

\subsection{Experiment settings} 
Following the previous work~\cite{xie2023disentangled}, we employ signals transmitted by 59 TI CC2530 ZigBee devices and captured by a USRP N210 receiver at various locations. All ZigBee transmitters operate in the 2.4 GHz band with a maximum output power of 19 dBm. The receiver is configured with a sampling rate of 10 MSample/s, resulting in each preamble signal $\vecx$ consisting of $M=1280$ samples. The signals are segmented into eight distinct subsets according to their characteristics. As shown in Table~\ref{tb:dataset}, the training and validation sets, acquired at a signal-to-noise ratio~(SNR) of 30 dB, are employed to derive the base model. Test sets T1–T3 contain signals from previously unseen devices or those subjected to aging effects, while test sets U1–U3 comprise signal data from unseen devices collected under previously unobserved channel conditions. All test sets were acquired under SNR $\approx$ 20 dB. 

For each test set, we allocate 20\% of the data to adapt the base model using different methods, while the remaining 80\% is reserved for evaluating the effectiveness of the adaptation approach. We report the average performance across all test sets. Since RLA combines multiple LoRAs, when assessing a specific test set, we use the LoRAs adapted from the 80\% held-out portion of other test sets as building blocks to evaluate performance.

\bflag{
In the LoRA module pretraining stage, each module within the RLA framework is pretrained on a specific subset of channel conditions. The training process runs for a minimum of 150 epochs with a learning rate of 0.01 and terminates once the AUC reaches 0.99. After pretraining, the LoRA modules are frozen and stored in a module pool. During inference, the RLA framework dynamically selects and combines these pretrained modules through weighted integration, without requiring any additional weight updates or fine-tuning.
}

\bflag{
In the CMA-ES optimization phase, we set the hyperparameters following the literature~\cite{hansen2016cma}. Specifically, we adopted the hyperparameter settings from the literature~\cite{hansen2016cma}: the population size $\lambda$ determines the exploration breadth in each generation and is set to $\lambda = 4 + \lfloor 3 \times \log{K} \rfloor$. The number of parents $\mu$ is set to $\mu = \lambda/2$ and serves as the consensus for guiding the search direction. The initial step size $\sigma_0 = 0.7$ represents the initial search radius, and the maximum number of iterations is set to 20.
}

To ensure a fair comparison, we employ the same neural network architecture as in~\cite{xie2023disentangled} for all baseline methods and use the SGD optimizer with momentum $m=0.9$ and learning rate $\eta=0.01$ consistently across all channel adaptation methods.
\begin{table*}[htbp]  

\caption{\bflag{Analysis of the hyperparameters $r$ and $K$ in the proposed RLA.}}

\centering
\begin{threeparttable}[b]
\begin{tabular}{lccccc}  

\toprule   
\bflag{$r=$} & \bflag{$1$} & \bflag{$2$} & \bflag{$4$} & \bflag{$8$} & \bflag{N/A}  \\
\midrule
\bflag{AUC(\%)} & \bflag{99.05 $\pm$ 1.08} & \bflag{99.15 $\pm$ 0.91} & \bflag{\textbf{99.19 $\pm$ 0.90}} & \bflag{99.09 $\pm$ 0.94} & \bflag{N/A} \\
\bottomrule
\toprule
\bflag{$K=$} & \bflag{$1$} & \bflag{$2$} & \bflag{$3$} & \bflag{$4$} & \bflag{$5$} \\
\midrule
\bflag{AUC(\%)} & \bflag{98.88 $\pm$ 1.00} & \bflag{98.95 $\pm$ 1.02} & \bflag{99.16 $\pm$ 0.82} & \bflag{99.18 $\pm$ 0.89} & \bflag{\textbf{99.19 $\pm $ 0.90}} \\
\bottomrule

\end{tabular}
\end{threeparttable}
\label{tb:hp}
\end{table*}
\bflag{\subsection{Hyperparameter Analysis}}

\bflag{
To investigate the impact of the LoRA rank $r$ and the number of modules $K$ on the performance of the proposed RLA, we conducted preliminary studies comparing different configurations: (1) varying rank $r \in \{1, 2, 4, 8\}$ with $K = 5$ fixed, and (2) varying the number of modules $K \in \{1, 2, 3, 4, 5\}$ with $r = 4$ fixed. Each experiment was repeated 10 times under identical conditions, using the same set of LoRA modules of varying ranks and consistent parameter settings for CMA-ES. The results are summarized in Table \ref{tb:hp}.
}

\bflag{
As presented in Table \ref{tb:hp}, the model achieves peak performance on the validation set when $r = 4$, whereas performance declines at $r = 8$. This indicates that, for the single-channel scenario, a LoRA module with $r = 4$ is sufficient to capture its discriminative features, while an excessively high rank may lead to overfitting on noise, thereby reducing model performance. 
}

\bflag{
Regarding the variation in the number of modules, $K$, we similarly observed that as known channels accumulate, a larger number of LoRA modules leads to better model performance in RLA. Since our dataset comprises at most six wireless channel types, with one reserved as the test set, we set $K=5$ to evaluate the model’s performance.
}

\subsection{Performance comparisons}

\begin{figure*}[htbp]
    \setlength{\textfloatsep}{3pt}
    \centering
    \begin{minipage}[h]{0.64\linewidth}
        \centering
        \captionof{table}{Experimental results across different baseline methods.}
        \resizebox{\linewidth}{!}{
\centering
\begin{threeparttable}[b]
\begin{tabular}{l|l|c|c|c}
\toprule
\multirow{2}*{Baselines} & \multirow{2}*{Adaptation method} & \multirow{2}*{Base Model} & \multicolumn{2}{c}{Performance} \\
\cline{4-5}
& & & AUC(\%) & EER(\%)\\
\bottomrule
SSL-FT~\cite{shen2024federated} & Full fine-tuning  & Self-supervised & $ 85.73 \pm 10.71$ & $ 20.63 \pm 11.14 $ \\
\bottomrule
ML-FT\cite{huan2025LLRF} & Full fine-tuning & \multirow{4}*{ML-RFF~\cite{xie2021generalizable}} & $ 81.76 \pm 13.50 $ & $ 23.28 \pm 14.08$ \\ 
\cline{1-2}
\cline{4-5}
ML-LoRA-2 & LoRA~($r$ = 2)~\cite{hu2022lora} &\multirow{4}*{(AUC=97.11\%)} & $ 98.15 \pm 2.03 $ & $ 5.90 \pm 3.39 $ \\
\cline{1-2}
\cline{4-5}
ML-LoRA-4 & LoRA~($r$ = 4) & \multirow{4}*{(EER=8.22\%)} & $ 97.73 \pm 2.71 $ & $ 6.71 \pm 4.21 $ \\
\cline{1-2}
\cline{4-5}
ML-LoRA-8 & LoRA~($r$ = 8) & & $ 96.99 \pm 3.56 $ & $ 8.05 \pm 5.19 $ \\
\cline{1-2}
\cline{4-5}
ML-RLA\dag & RLA~($r$=4, $K$=5) & & \textbf{98.94 $\pm$ 0.88} & \textbf{4.33 $\pm$ 2.33} \\
\bottomrule
DR-FT & Full fine-tuning & \multirow{4}*{DR-RFF~\cite{xie2023disentangled}} & $96.61 \pm 4.05$  & $ 8.36 \pm 4.47 $ \\ 
\cline{1-2}
\cline{4-5}
DR-LoRA-2 & LoRA~($r$ = 2) & \multirow{4}*{(AUC=98.87\%)}& $ 98.68 \pm 1.59 $ & $ 4.60 \pm 3.47 $ \\
\cline{1-2}
\cline{4-5}
DR-LoRA-4 & LoRA~($r$ = 4) & \multirow{4}*{(EER=3.92\%)}& $ 98.11 \pm 2.82 $ & $ 5.79 \pm 4.44 $ \\
\cline{1-2}
\cline{4-5}
DR-LoRA-8 & LoRA~($r$ = 8) & & $ 97.83 \pm 2.70 $ & $ 6.52 \pm 4.13 $ \\
\cline{1-2}
\cline{4-5}
DR-RLA\dag & RLA~($r$=4, $K$=5) & & \textbf{99.19 $\pm$ 0.90} & \textbf{3.32 $\pm$ 2.88} \\
\bottomrule
\end{tabular}
\begin{tablenotes}
\item[\dag]Proposed method in this paper. 
\end{tablenotes}
\end{threeparttable}
}
        \label{tb:result}
    \end{minipage}
    \hfill
    \begin{minipage}[h]{0.35\linewidth}
        \centering
        \includegraphics[width=\linewidth]{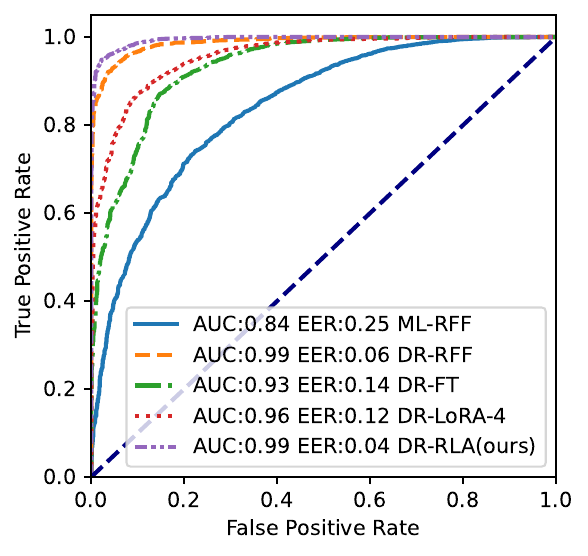}
        \captionof{figure}{ROC curves of baselines on U3.}
        \label{fig:roc}
    \end{minipage}
\end{figure*}

\begin{figure}
    \centering
    \includegraphics[width=\linewidth]{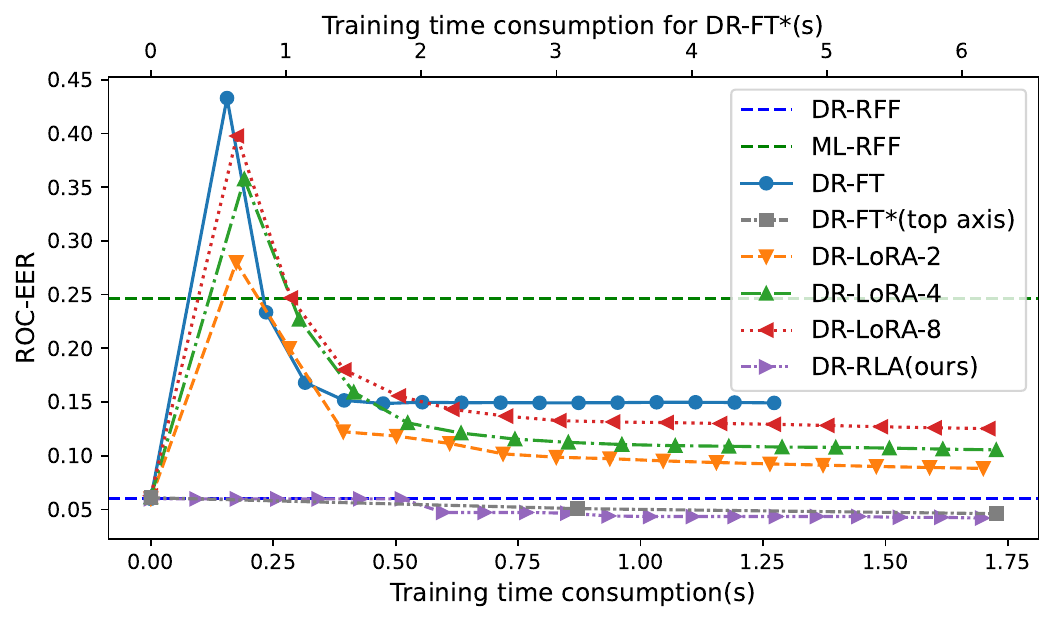}
    \caption{Learning curves of baselines on U3.}
    \label{fig:eers}
\end{figure}

To verify the effectiveness of the proposed RLA, we evaluate proposed RLA comparing the performance of the RFFs with that of the baseline algorithms under different base model settings, i.e., ML-RFF~\cite{xie2021generalizable} and DR-RFF~\cite{xie2023disentangled}. Evaluation results are presented in Table~\ref{tb:result}, Fig.\ref{fig:roc} and Fig.\ref{fig:eers}. 

Overall, compared to other approaches, the proposed RLA achieves effective transfer to new devices and channels with low computational and time overhead, requiring only a small fraction, i.e.,20\%, of the target test set. As demonstrated in Table~\ref{tb:result}, when ML-RFF is used as the base model, all adaptive methods outperform the base model significantly, with RLA delivering the most substantial performance gain. In contrast, SSL-based methods struggle to learn effective initial representations during pre-training, resulting in poor adaptation to new channels or devices when limited test data is available.

On the other hand, when DR-RFF is used as the base model, all adaptation methods, except the proposed RLA, exhibit significant performance degradation. The DR-RFF employs adversarial regularization to mitigate overfitting to the training data. However, FT and LoRA-based methods still optimize under the standard MLE objective, which creates a significant misalignment with the feature representations utilized by DR-RFF, ultimately impairing performance. In contrast, the proposed RLA method restricts the optimization space by reducing the number of trainable LoRA combination parameters. This constrained adaptation achieves a favorable trade-off, enabling consistent performance gains even with limited data, while minimizing feature distortion introduced by integrating multiple LoRA modules into the base model. Moreover, as illustrated in Fig.\ref{fig:eers}, the sampling-based optimization algorithm converges faster than LoRA, achieving the second-fastest convergence speed after FT. Notably, even when compared to the baseline DR-FT* in Fig.\ref{fig:eers}, which fine-tuned with the same training data as the LoRA models within RLA, the proposed RLA requires only 1/6 of the training time to achieve comparable performance.

These experimental results demonstrate the superior performance of the proposed RLA, enabling performance gains using only a few signal samples in unknown environments or with unknown devices. It provides a short-term, efficient adaptive solution for RFF applications under rapidly changing environmental conditions.

\section{Conclusion}
In this paper, we propose a lightweight and efficient adaption method for RFF extraction based on LoRA, addressing the critical challenges of open-set device authentication in dynamic and heterogeneous wireless environments. By pretraining LoRA modules for different propagation environments and dynamically combining them at runtime, our approach achieves superior adaptability with significantly reduced computational overhead compared to full fine-tuning and non-fine-tuning methods. Experimental results on a real-world testbed demonstrate the effectiveness of the proposed method, achieving comparable performance with an 83\% reduction in training time. The proposed solution offers a practical, scalable, and energy-efficient framework for robust physical-layer authentication in real-world IoT applications, paving the way for more reliable and deployable secure wireless systems.

\ifCLASSOPTIONcaptionsoff
  \newpage
\fi

\bibliographystyle{./IEEEtran}

\bibliography{IEEEabrv, references}

@Article{wang2016wireless,
  title={Wireless physical-layer identification: Modeling and validation},
  author={Wang, Wenhao and Sun, Zhi and Piao, Sixu and Zhu, Bocheng and Ren, Kui},
  journal={{IEEE} Trans. Inf. Forensics Secur.},
  volume={11},
  number={9},
  pages={2091--2106},
  year={Apr. 2016},
  publisher={IEEE}
}

@Article{hou2014physical,
  title={Physical layer authentication for mobile systems with time-varying carrier frequency offsets},
  author={Hou, Weikun and Wang, Xianbin and Chouinard, Jean-Yves and Refaey, Ahmed},
  journal={{IEEE} Trans. Commun.},
  volume={62},
  number={5},
  pages={1658--1667},
  year={Apr. 2014},
  publisher={IEEE}
}

@Article{xie2024radio,
  title={Radio frequency fingerprint identification for Internet of Things: A survey},
  author={Xie, Lingnan and Peng, Linning and Zhang, Junqing and Hu, Aiqun},
  journal={Secur. Saf.},
  volume={3},
  pages={2023022},
  year={2024},
  publisher={EDP Sciences and CSPM}
}

@Article{danev2012physical,
  title={On physical-layer identification of wireless devices},
  author={Danev, Boris and Zanetti, Davide and Capkun, Srdjan},
  journal={{ACM} Comput. Surv.},
  volume={45},
  number={1},
  pages={1--29},
  year={Dec. 2012},
  publisher={ACM New York, NY, USA}
}

@Article{merchant2018deep,
  title={Deep learning for {RF} device fingerprinting in cognitive communication networks},
  author={Merchant, Kevin and Revay, Shauna and Stantchev, George and Nousain, Bryan},
  journal={{IEEE} J. Sel. Top. Signal Process.},
  volume={12},
  number={1},
  pages={160--167},
  year={Jan. 2018},
  publisher={IEEE}
}

@Article{xie2021generalizable,
  title={A generalizable model-and-data driven approach for open-set {RFF} authentication},
  author={Xie, Renjie and Xu, Wei and Chen, Yanzhi and Yu, Jiabao and Hu, Aiqun and Ng, Derrick Wing Kwan and Swindlehurst, A Lee},
  journal={{IEEE} Trans. Inf. Forensics Secur.},
  volume={16},
  pages={4435--4450},
  year={Aug. 2021},
  publisher={IEEE}
}

@Article{xie2023disentangled,
  title={Disentangled representation learning for {RF} fingerprint extraction under unknown channel statistics},
  author={Xie, Renjie and Xu, Wei and Yu, Jiabao and Hu, Aiqun and Ng, Derrick Wing Kwan and Swindlehurst, A Lee},
  journal={{IEEE} Trans. Commun.},
  volume={71},
  number={7},
  pages={3946--3962},
  year={Apr. 2023},
  publisher={IEEE}
}

@ARTICLE{zhang2024real,
  author={Zhang, Zechen and Li, Guyue and Shi, Jitong and Li, Haobo and Hu, Aiqun},
  journal={{IEEE} Trans. Veh. Technol}, 
  title={Real-world aircraft recognition based on {RF} fingerprinting with few labeled {ADS-B} signals}, 
  year={Feb. 2024},
  volume={73},
  number={2},
  pages={2866--2871}
}

@ARTICLE{zhang2024DFLNet,
  author={Zhang, Tiantian and Xu, Dongyang and Ren, Pinyi and Yu, Keping and Guizani, Mohsen},
  journal={{IEEE} Trans. Veh. Technol}, 
  title={{DFLNet}: Deep federated learning network with privacy preserving for vehicular {LoRa} nodes fingerprinting}, 
  year={Feb. 2024},
  volume={73},
  number={2},
  pages={2901--2905},
}

@ARTICLE{huan2025LLRF,
  author={Huan, Xintao and Wu, Changfan and Lei, Yulu and Liu, Jiamin and Hao, Yi and Wang, Jianchao},
  journal={{IEEE} Trans. Veh. Technol}, 
  title={{LLRF}: Towards long-term {LoRa} radio frequency fingerprint identification based on transfer learning}, 
  year={Jul. 2025},
  volume={1},
  number={1},
  pages={1--6},
}

@InProceedings{hu2022lora,
  author       = {Hu, Edward J. and Shen, Yelong and Wallis, Phillip and Allen{-}Zhu, Zeyuan and Li, Yuanzhi and Wang, Shean and Wang, Lu and Chen, Weizhu},
  title        = {{LoRA}: Low-Rank adaptation of large language models},
  booktitle    = {Proc. Int. Conf. Learn. Representations},
  year         = {Apr. 2022},
  publisher    = {OpenReview.net}
}

@Article{del2024fingerprint,
  title={Fingerprint extraction through distortion reconstruction ({FEDR}): A {CNN}-based approach to {RF} fingerprinting},
  author={del Arroyo, Jose A Gutierrez and Borghetti, Brett J and Temple, Michael A},
  journal={{IEEE} Trans. Inf. Forensics Secur.},
  volume={19},
  pages={9258--9269},
  year={Sep. 2024},
  publisher={IEEE}
}

@Article{jagannath2022comprehensive,
  title={A comprehensive survey on radio frequency ({RF}) fingerprinting: Traditional approaches, deep learning, and open challenges},
  author={Jagannath, Anu and Jagannath, Jithin and Kumar, Prem Sagar Pattanshetty Vasanth},
  journal={Comput. Networks},
  volume={219},
  pages={109455},
  year={Dec. 2022},
  publisher={Elsevier}
}

@Article{shen2024federated,
  title={Federated radio frequency fingerprint identification powered by unsupervised contrastive learning},
  author={Shen, Guanxiong and Zhang, Junqing and Wang, Xuyu and Mao, Shiwen},
  journal={{IEEE} Trans. Inf. Forensics Secur.},
  volume={19},
  pages={9204--9215},
  year={Sep. 2024},
  publisher={IEEE}
}

@Article{peng2023supervised,
  title={Supervised contrastive learning for {RFF} identification with limited samples},
  author={Peng, Yang and Hou, Changbo and Zhang, Yibin and Lin, Yun and Gui, Guan and Gacanin, Haris and Mao, Shiwen and Adachi, Fumiyuki},
  journal={{IEEE} Internet Things J.},
  volume={10},
  number={19},
  pages={17293--17306},
  year={May 2023},
  publisher={IEEE}
}

@InProceedings{iyiparlakouglu2024optimizing,
  title={Optimizing radio frequency fingerprinting for device classification: A study towards lightweight {DL} models},
  author={{\.I}yiparlako{\u{g}}lu, Raif and Awan, Maaz Ali and Dalveren, Yaser and Kara, Ali},
  booktitle={Proc. Int. Conf. Commun., Signal Process., Appl.},
  pages={1--6},
  address={Istanbul, Turkiye},
  year={Dec. 2024},
  organization={IEEE}
}

@Article{hanna2021open,
  title={Open set wireless transmitter authorization: Deep learning approaches and dataset considerations},
  author={Hanna, Samer and Karunaratne, Samurdhi and Cabric, Danijela},
  journal={{IEEE} Trans. Cogn. Commun. Netw.},
  volume={7},
  number={1},
  pages={59--72},
  year={Dec. 2021},
  publisher={IEEE}
}

@Article{yu2019robust,
  title={A robust {RF} fingerprinting approach using multisampling convolutional neural network},
  author={Yu, Jiabao and Hu, Aiqun and Li, Guyue and Peng, Linning},
  journal={{IEEE} Internet Things J.},
  volume={6},
  number={4},
  pages={6786--6799},
  year={Apr. 2019},
  publisher={IEEE}
}

@Article{shen2022towards,
  title={Towards scalable and channel-robust radio frequency fingerprint identification for {LoRa}},
  author={Shen, Guanxiong and Zhang, Junqing and Marshall, Alan and Cavallaro, Joseph R},
  journal={{IEEE} Trans. Inf. Forensics Secur.},
  volume={17},
  pages={774--787},
  year={Feb. 2022},
  publisher={IEEE}
}

@Article{hansen2016cma,
  author       = {Nikolaus Hansen},
  title        = {The {CMA} Evolution Strategy: {A} Tutorial},
  journal      = {CoRR},
  volume       = {abs/1604.00772},
  year         = {Apr. 2016},
  eprinttype   = {arXiv},
  eprint       = {1604.00772},
  bibsource    = {dblp computer science bibliography, https://dblp.org},
}

@InProceedings{shazeer2017moe,
  author       = {
      Shazeer, Noam and
      Mirhoseini, Azalia and
      Maziarz, Krzysztof and
      Davis, Andy and
      V. Le, Quoc and
      E. Hinton, Geoffrey  and
      Dean, Jeff
  },
  title        = {\bflag{Outrageously large neural networks: The sparsely-gated mixture-of-experts layer}},
  booktitle    = {\bflag{Proc. Int. Conf. Learn. Representations}},
  publisher    = {\bflag{OpenReview.net}},
  year         = {\bflag{Apr. 2017}},
}

\end{document}